# Hypergraphic partitioning of quantum circuits for distributed quantum computing

Waldemir Cambiucci, Regina Melo Silveira, Wilson Vicente Ruggiero

Department of Computing Engineering and Digital Systems
University of Sao Paulo, SP, Brazil
(Updated: February 3, 2023)

**Abstract** - In the context of NISQ computers - Noise Intermediate Scale Quantum, it is a consensus that the distribution of circuits among processing agents is a viable approach to get greater scalability with small machines. This approach can increase the combined computational power at the cost of dedicating qubits to get the communication between partitions. Then comes the challenge of reducing this cost of communication-based on efficient circuit partitioning strategies. In the literature, we find numerous works exploring different partitioning approaches that need a consensus on the best method for distribution scenarios. In this context, this work aims to reduce the consumption of communication qubits generated during the physical partitioning of quantum circuits in the distributed quantum computing scenario. We present a new method for partitioning quantum circuits in a hypergraphic representation. The proposed approach uses a heuristic partitioning algorithm on the hypergraphic representation to reduce the number of communication qubits between the partitions, generating a lower cost of communication in distributed quantum systems. For this, we follow a three-step approach: first, a logical segmentation and grouping of gates for optimization and reuse of communication qubits between future partitions, considering the dimensions of width (total qubits), size (total of gates), and depth (total time steps for execution) of the circuit to be distributed; second, a translation of the quantum circuit into a hypergraphic representation, with vertices and edges representing qubits and multi-qubit gates, respectively; finally, use of a variation of the FIDUCCIA-MATTHEYSES heuristic method of hypergraphic partitioning, with recursive bipartite and direct k-way approaches, to generate partitions in balanced and unbalanced scenarios of distributed quantum systems. With this approach, we obtained partial results with a more than 50% reduction in the communication cost generated for the bipartite partitioning against a partitioning done randomly on benchmark circuits. The expectation is that future experiments with multiple partitions will continue with equal or better results than previous works, supporting the evolutionary process of realizing distributed quantum computing with machines in heterogeneous processing and communication scenarios.

## 1 – Introduction

**Motivation** – today, it is a consensus that the distribution of circuits in several small processing units is a viable approach to obtain greater scalability with NISQ computers, increasing the power of quantum computers in combination, as mentioned by LOMONACO [3].

Hundreds or thousands of qubits are estimated as necessary for several application scenarios for quantum to obtain the expected quantum advantage over the solutions present in classical computing. Complex circuits that require a considerable number of qubits still cannot be executed on available machines, as this offer just a few hundred qubits for processing. Still, with numerous challenges related to gate reliability, high rates of decoherence and scalability challenges, and deployment of quantum error correction [4][5], we must find new ways to improve scalability in quantum computing. While numerous laboratories and manufacturers seek to develop quantum computers with a more significant number of qubits and, consequently, more processing capacity [7][8], we find ample space for research and exploration of solutions for distributed quantum computing among smaller processing units.

In this work, we focus on the partitioning of quantum circuits for distributed quantum computing, and we present a new method for the partitioning of quantum circuits in hypergraphic representation to minimize the communication between partitions, at the same time supporting configurable scenarios according to the capabilities and constraints of processing units involved in distribution.

The partitioning of a circuit generates a cut or more in groups of binary gates, creating segments with different numbers of qubits that will be executed in different processing agents. Each partition will contain data qubits (used for processing) and communication qubits dedicated to exchanging information between partitions. A partitioning is considered balanced when the generated partitions have the same number of qubits. The unbalanced scenario contemplates partitions with different amounts of qubits, representing heterogeneous infrastructures of quantum machines. The big challenge in this context is to reduce the need to consume communication qubits, prioritizing qubits for processing and avoiding production costs for these links between partitions, performed through quantum teleportation protocols.

The method proposed in this work is based on the translation of the input quantum circuit into a hypergraphic representation following the partitioning process with a variation of the algorithm of FIDUCCIA-MATTHEYSES (FM) [10]. As a preparation for the input circuit before its translation into hypergraphic representation, we looked for specific patterns of binary gates and grouping of nearby logic gates to reduce the consumption of communication qubits during partitioning. These reuse patterns generate the so-called grouping vertices, added to the generated hypergraph, guiding future cut points by the partitioning heuristic with the FM method.

Thus, we follow three steps for the partitioning of quantum circuits in hypergraphic representation in a balanced and unbalanced way:

• *logical segmentation and grouping of gates*, *considering the recognition of clustering patterns of nearby gates. We also evaluated the impact of circuit segmentation based on input circuit dimensions, such as width (total qubits), size (total gates), and depth (total time steps);*

• *translation of the quantum circuit into a hypergraphic representation*, *with a conversion algorithm for creating vertices and edges representing qubits and gates of multiple qubits, respectively;*

• *Finally, the use of a* **variation of the FIDUCCIA-MATTHEYSES method of hypergraphic partitioning**, *with recursive bipartite and direct k-way approaches. By doing, we can generate partitions in balanced and unbalanced scenarios of distributed quantum systems.*

The generated sub-hypergraphics can be translated into quantum circuits, ready to be submitted for execution in distributed QPUs.

Once based on an extension of the FM algorithm, the partitioning method proposed in this work becomes faster and more efficient for unbalanced partitioning scenarios, with direct application to currently distributed quantum computing environments with heterogeneous machines. In scenarios of low availability of quantum devices and a limited number of qubits, the distributed system with heterogeneous machines is a crucial solution to increase the quantum processing capacity. As recognized by PRESKILL [1], this is an important stage before realizing fault-tolerant and more scalable quantum computing, the basis for the future quantum internet [6].

The following figure illustrates the sequence of steps proposed in this work.

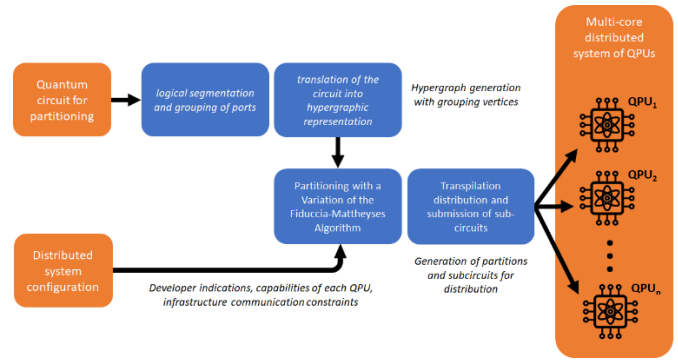

*Figure 1 – Steps of the partitioning process of quantum circuits, designed and used in this work.*

**This Paper -** this research aims to propose a new approach for the distribution of quantum circuits in multiple QPUs, through the segmentation and hypergraphic representation of quantum circuits, with partitioning performed with a variation of the heuristic algorithm of FIDUCCIA-MATTHEYSES, generating partitions with lower costs of Communication.

As specific objectives of this work, we will:

• Propose an approach for the logical optimization, segmentation, and gate grouping of an input quantum circuit, considering the recognition of ebits reuse patterns and the dimensions of width, size, and depth of the input circuit and impact on the segmentation;

• Propose a variation of the FIDUCCIA-MATTHEYSES algorithm, with recursive bipartite approaches and direct k-way partitioning, to generate k partitions representing different QPUs of balanced and unbalanced distributed systems;

• Identify a correlation between the dimensions of the input quantum circuit and the results of partitioning with the consumption of communication qubits between partitions, guiding possible development patterns and optimal patterns for distributed circuits.

• Consolidate a development framework for the different stages of this project. For example, we have tools for intermediate representation, translation to hypergraphic representation, hypergraphic partitioning, quantum simulation, and circuit submission on quantum platforms.

**Contributions -** The discussion about partitioning quantum circuits for heterogeneous distribution environments, with reduced consumption of communication qubits, is a relevant and strategic component in realizing distributed quantum computing. Thus, as main contributions generated by this research, we mention:

- Propose a new partitioning technique for quantum circuits with gate segmentation and grouping for non-local operators, using a hypergraphic approach and a variation of the Fiduccia-Matheyses algorithm;

- Dynamic quantum circuit design patterns and compile-time synthesis for better partitioning results in multicore distributed systems;

- A reference architecture for distributed quantum computing, considering the partitioning and distribution of circuits in heterogeneous hardware scenarios, using as a basis the quantum computer architecture design proposed by BANDIC et al. [8];

- A development framework for partitioning quantum circuits, representation, and analysis by hypergraphs, submission of circuits to real quantum machines and simulators, and tools for analyzing results, published through the GitHub repository.

## 2 – Problem Definition

We must consider different resource availability scenarios for the circuit partitioning challenge in heterogeneous environments. As presented by FERRARI [44], the communication challenge between QPUs involves the dedication of communication qubits, reducing the computational power offered with data qubits for the execution of circuits.

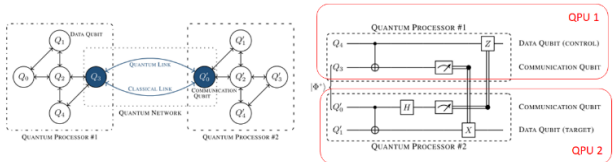

Figure 2 – Dedicated qubits for communication between QPUs, an expensive resource to be minimized when processing is distributed across multiple QPUs. Source: FERRARI [44].

In distributed scenarios with multiple QPU processing units, the challenge of partitioning with reduced communication costs is vital due to the impact on the processing power available for the generated partitions.

An approach conceived and used here considers the sum of qubits offered by the network processing units as a limit for the size of the input quantum circuit that will be distributed.

Thus, we define the following environment variables:

- $n$ = number of qubits in the input circuit
- $k$ = number of participating QPUs
- $w$ = number of partitions resulting from partitioning
- $q_i$ = number of qubits per participating QPU, where $i = 1..k$
- $S$ = total qubits of participating QPUs, where $S = \sum q_i$
- $e$ = number of communication qubits per resulting partition
- $o$ = number of operations grouped by the resulting partition
- $r$ = relation between the number of ebits and operations per partition, $r = e/o$

With these variables in focus, we can work with two different scenarios of partitioning and distribution of circuits between quantum processing units connected in a network or as part of a multicore machine:

- **Balanced scenario:** for a quantum circuit qc with n qubits and distributed environment of k QPUs, partition the circuit into w partitions, where $S >= n$. We consider here a scenario of QPUs with equal numbers of qubits ($q_1 = q_2 = q_3 = q_k$); in this scenario, each partition will run on one QPU, where $w = k$.

- **Unbalanced scenario:** for a qc quantum circuit with n qubits and distributed environment of k QPUs, partition the circuit into w partitions, where $S >= n$. We consider here a scenario of QPUs with different numbers of qubits ($q_1 \neq q_2 \neq q_3 \neq q_k$); in this scenario, the number of partitions remains equal to the number of available QPUs, where $w = k$. Each generated partition can have a different number of qubits, between data and communication qubits, according to the configuration of each QPU.

So, for an input quantum circuit with n qubits and a distributed system with k QPUs, we must partition the circuit, generating w partitions, each partition $w_i$ being destined for a different QPU $k_i$, reducing the number of ebits needed for communication between the partitions.

The $w > k$ scenario will not be the focus of this work when the number of partitions generated is greater than the number of QPUs available on the network. In this situation, we must consider scheduling the partitions between the QPUs present, and it is expected that more than one partition will run on the same QPU. We will leave this scenario to future research and explorations.

We are exploring an efficient method of partitioning input quantum circuits into smaller partitions intended for execution on NISQ machines with different amounts of qubits available for communication and data.

For the two focus scenarios (balanced and unbalanced), the challenge is to determine the ratio r for the resulting partitioning, that is, the ratio between the number of qubits dedicated to communication between partitions and the number of operations in the total of each partition. This ratio becomes an important target for reduction, indicating that the proportion of resources dedicated to communication infrastructure is smaller, prioritizing qubits for the quantum processing of the algorithm. Circuit partitioning should therefore consider different initial configurations of the distributed environment, better guiding the partitioning solution.

Distributed quantum computing involves numerous aspects and elements, such as communication protocols, software and hardware layers, resource allocation tools, and native challenges of current technologies used in constructing quantum computers and implementing qubits. Thus, for the partitioning problem of quantum circuits, it is also essential to define the boundaries and target scenarios of research for the correct extrapolation of results and application of algorithms.

This research does not determine hardware aspects or specific communication protocols as part of the communication between partitions. In this way, we use abstractions implemented in *OpenQASM* intermediate representation [65], encapsulating calls of communication primitives between partitions. This abstraction offers the advantage of supporting both quantum network simulation environments and future scenarios of multicore machines, foreseen for the near future. Quantum computing environments distributed in a local simulator, such as the *QuNetSim* [62] or *NETSQUID* [63] platforms, are suitable examples for studying quantum computer networks by discrete events and simulated via classical communication. Using this approach, each partition generated in the circuit distribution process can use the function encapsulation feature of OpenQASM [65], including sync points for communications equivalent to *CAT-ENTANGLER* and *CAT-DISENTANGLER* [3]. The translation in different platforms can be easily done, as the user functions from an intermediate representation.

Finally, this work considers the scenario of machines with multiple cores, where from the same access subscription, we can indicate the desired backends within the same platform. To exercise this scenario, we will use resources from the QISKIT platform [66] from IBM Quantum and resources from MICROSOFT AZURE Quantum [130] [131], which allows the encapsulation of multiple circuits for the same quantum computer, and the execution of multiple circuits for different backends from the same module executed in Python language. These scenarios simulate the context of coordination subsystems and triggering of partitioned circuits, as foreseen for real distributed environments with different technologies of hardware from differente providers. Still, this empirical and practical context offers excellent benefits in generating data and results for efficiency analysis and evaluation of the partitioning algorithms experimented with in this work.

It is essential to point out that the partitioning approach for multiple cores differs from the distributed approach on multiple machines of different platforms. Other layers of software and communication must be considered in the context of multiple machines from different vendors [67]. In this case, we will have the requirement for different access licenses and subscriptions between manufacturers, which is an example of quantum computing as a service with multi-vendors.

Thus, this research actively works on the intermediate representation of quantum circuits, choosing the *OpenQASM* representation as a basis. It considers the future of cross-platform communication between quantum machines from different manufacturers, facilitating the translation process, management, and exchange of circuits between different environments.

### 3 – Survey on Distributed Quantum Computing

During the literature exploration and systematic literature review, numerous works related to distributed quantum computing and partitioning quantum circuits were found. Thus, we present below a brief history that surveys the principal outstanding works supporting this research and the proposed extensions.

Since the first discussions on quantum communication and distributed quantum computing were present, we have noticed a diversity of approaches relating to challenge such as stacks of protocols and communication, local and remote quantum gates, communication costs, mechanisms for generating quantum links and classical channels of communication, among others [14]. A key feature of many of the quantum protocols and non-local quantum gates for communication between distributed quantum systems is quantum entanglement, where entangled quantum states are shared by qubits persisted on different nodes in a distributed system. This procedure is costly for current NISQ computers since it takes away processing power, dedicating qubits for communication and links between nodes.

In the last 20 years, we found several works evolving the themes related to the construction of distributed quantum computing, always with analogies to the concepts and approaches we saw in the classic distributed computing environment, with glimpses for the future quantum internet and its applications [15]. Following this evolution, we see a brief history of the main works found below.

The first ideas on distributed quantum computing can be associated with GROVER [16] and his work in 1997 on algorithm distribution for partial search, segmenting data blocks between partitions. In the same year, CLEVE and BUHRMAN [17] stimulated a discussion on quantum communication using quantum entanglement, which is fundamental for current quantum communication protocols. In 1999, CIRAC et al. [18] presented a vital discussion about distributed quantum computation in noisy channels, initiating the debate about communication protocols and decoherence during quantum communication.

In 2001, YEPEZ [19] made a significant contribution to the classification of quantum computers for distributed systems into two categories, suggesting that some complex classical problems could be solved in hybrid quantum computers, formed essentially by small quantum machines connected through classical channels of information. Thus, he defined Type I for quantum computers that use quantum communication between subsystems. Each qubit can be entangled with the number of qubits participating in a network for this type of machine. The quantum computer exploits the classical communication between distributed network subsystems for type II machines, allowing multiple quantum systems to connect through classical communication channels. This approach is still used today for several quantum network simulation frameworks as an essential tool for developing communication protocols and connection primitives.

In the following years, two other works deserve mention in the context of communication primitives for distributed quantum computing: In 2004, YIMSIRIWATTANA [20] and LOMONACO [3] published articles proposing communication primitives CAT-ENTANGLER and CAT-DISENTANGLER that were widely used in the literature and are considered here as fundamental for the study of communication between partitions. Despite calculating the number of communication gates between partitions, the works do not deal with the reduction of communication costs generated by the distribution.

In 2008, VAN METER et al. [21] presented a distributed quantum circuit for the full adder. In this work, the author split the algorithm into two separate quantum circuits and then communicated with each other through teleportation channels. No approach was presented in this work to reduce the cost of communication generated, being interesting the approach adopted for dividing the circuit into two parts.

In 2009, FENG [22] presented an algebraic language for modeling quantum circuits in the context of distributed quantum computing, which was widely referenced for its formalism and efficiency in the characterization of quantum circuits. During this research, we will use concepts and foundations inspired by FENG.

In 2012, STRELTSOV et al. [23] proposed a form of distributed entanglement and provided the minimum quantum cost to send an entangled composite state over a long distance. It was the first work to discuss communication costs and the impact of distributed communication on quantum entanglement.

In 2013, BEALS et al. [24] presented a hypercube graphical approach to represent a quantum distributed system. This representation allowed the emulation of a simple quantum circuit in the form of a distributed system through nodes connected by the hypercube. Again, no discussion was made about reducing communication costs between participants in this work.

In 2016, the work by VAN METER and DEVITT [25] was considered a milestone for distributed quantum computing due to the segmented approach of computer generations and the impact obtained with the evolution of applied technologies. It is one of the first works to describe aspects of a quantum software stack, considering algorithm design, circuit design, hardware mapping, and topological optimizations, supporting some critical concepts about distributed quantum applications and quantum networks. In the same year, ZOMORODI et al. [26] presented an algorithm to optimize the number of qubits dedicated to communication in a distributed quantum circuit. In his approach, the algorithm processed the number of CNOT logic gates, treating different groupings of qubits directly on the vector representation of the circuit through the group G of the logic gates present.

Over the past few years, we have seen the continued development of quantum technologies, particularly in the hardware and communication disciplines. Observing the challenges related to the development of more powerful quantum computers, PRESKILL [1] created the concept of NISQ computers. In this approach, he categorized quantum machines by their limitations and challenges for the implementation of qubits, as well as the need for better techniques of error correction and noise reduction during quantum processing. In his article, he predicted that between 2018 and 2022, there would be continuous evolution in the correction of quantum errors and the different aspects of the construction of fault-tolerant quantum machines. The definition of NISQ machines increased the interest in exploring distributed quantum computing as an alternative solution for greater scalability in scenarios with multiple inferior machines working together to treat more complex algorithms.

In 2018, MARTINEZ and HEUNEN [27] inaugurated a new discussion on circuit partitioning and distributed quantum computing. They present a hypergraphic approach to the challenge of partitioning quantum circuits through a method based on two steps: the first, called pre-processing, optimizes equivalent gates; then, a hypergraphic partitioning is performed based on the KERNIGHAN-LIN algorithm [13], implemented through the KAHYPAR method [28] for the generation of comparison data with different types of circuits pre-selected for benchmarks, including QFT and Quantum Walk. This work inspired several recent initiatives, including using the same test circuits for comparison purposes.

In 2018, CALEFFI et al. [29] studied the challenges of designing the quantum internet. They also discussed the processing speed achieved using quantum computers connected via the quantum internet. In another work [30], the authors studied the creation of the quantum internet and considered teleportation as the primary protocol for transferring information.

Still, in 2018, MOGHADAM [31] explored techniques prioritizing the grouping of equivalent gates to save resources in partitioned circuits. ZULEHNER [32] also presented an relevant discussion about the transpilation and optimization of quantum circuits empirically for a market platform, IBM Quantum. Here, the authors addressed different aspects of optimization, mapping, and transpilation, which are important during steps associated with partitioning quantum circuits. It is worth remembering that transpilation refers to the process that combines stages of circuit translation and compilation for a given instruction set of a target quantum architecture.

More recently, several works have explored circuit partitioning through hypergraphic representation and hypergraphic partitioning algorithms of different types.

In 2020, DUNCAN [33] presented a technique for partitioning and simplifying quantum circuits using the graphical representation ZX-Calculus [34], which has been widely explored in several works, such as [35][36][37]. In the same year, DAVARZANI [38] made a new contribution to the study of circuit partitioning, extending the work of MARTINEZ [27] and comparing results with the same benchmark circuits used in previous work.

It is important to emphasize that the present research will follow the same benchmark circuits proposed by DAVARZANI

and MARTINEZ to allow future comparisons and calibration of results obtained with the FMEx method proposed here. DAEI et al. [39] also explored the partitioning of quantum circuits using an approach based on the KL algorithm [13], based on the continued application of the algorithm on each initial partition of the process. They used the MATLAB platform to validate results, with experiments based on the same synthesis used by BARENCO [40] and SHENDE [41].

Important considerations about the challenges of distributed quantum computing were presented by CUOMO [42], with particular attention to the treatment of multiple qubits involved in distributed systems and the operations related to quantum teleportation for the transfer of quantum information between devices. The author introduces the concept of TELEDATA, moving quantum information from one device to another through a teleportation process with entangled pairs of qubits. CHOU [43] is also referenced by the concept of a TELEGATE or remote quantum gate, indicating the formalism for binary gates between distributed partitions for a quantum system.

For a multicore quantum machine context, important aspects for constructing charge-distributed quantum compilers are presented by FERRARI et al. [44], which discusses data qubits and communication qubits, such as the need to dedicate at least one qubit in each processor to the communication operation, allowing the creation of a Bell state for consumption. The greater the number of CNOTs executed in parallel, taking advantage of the same generated Bell state channel, the lower the communication overhead during execution.

HU and KAIS [45] present an approach similar to previous works, dealing with the characterization of quantum circuits through configurations of functional qubits, decomposing any circuit into sequences of 1-qubit U unitary functions and CNOT binary gates. The work uses the characterization of circuits through functional qubits to guide a form of efficient circuit decomposition.

In 2021, SALEEM et al. [46] presented the QDCA – Quantum Divide and Conquer Algorithm to divide and conquer the circuit partitioning process for a distributed quantum architecture. As in previous works, the authors use the algorithm of KERNIGHAN-LIN [13] for partitioning the hypergraph representing the circuit for a scenario of balanced resources. The technique was validated for combinatorial optimization problems using the METIS platform [47]. In the same year, AVRON and ROZEN [48] applied the technique of decomposing circuits into Boolean functions and disjunctive normal forms (DNF), in the context of circuit distribution, with an insightful impact on some families of algorithms.

In 2022, WU et al. [49] created the AutoComm framework to identify burst communication concepts, a connectivity pattern between partitions that, when identified, allows the reduction of communication costs. AutoComm is presented as a framework for compilers, extracting patterns or burst patterns present in the input circuit before partitioning.

SUNDARAM et al. [50] presented a method called SPLIT algorithm for moving quantum teleportation points represented through the hypergraph model known in graph theory as ERDOS-RÉNYI [53]. To evaluate the results, the authors used random quantum circuits.

We thus saw a brief history of the last 20 years, on the development of distributed quantum computing, with a special focus on the challenge of partitioning and communication between subsystems. Over the years, several quantum algorithms have been developed with a focus on solving a wide variety of applications. Here we have examples such as phase estimation algorithms [54,55], Shor factorization algorithm [56,57], the algorithm for linear systems of HARROW-HASSIDIM-LLOYD [58], hybrid quantum algorithms, with consolidation or parameterization portions in classical computing [59] and more. Recently, we saw more research on quantum algorithms for machine learning [60], potentially applicable to a myriad of research and industry scenarios. These examples require a large computing capacity for current NISQ quantum computers, which underlies and motivates research around distributed quantum computing. Several other articles were published in the same period in related areas.

**4 – Reference architecture for distributed quantum computers**

From the reference architecture of quantum computers presented in previous works, the following figure uses an architecture of distributed quantum computers conceived and considered in this work for the layered organization of a distributed environment with n QPUs.

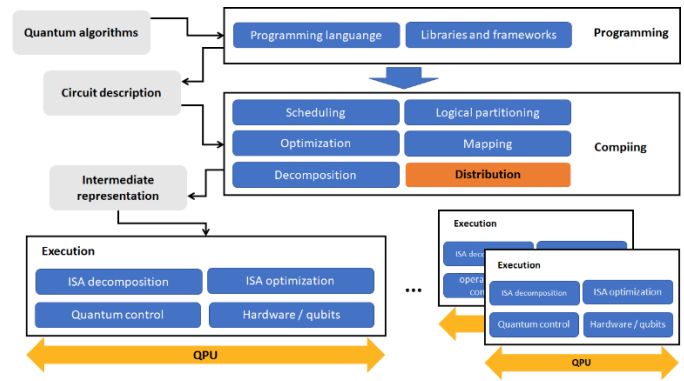

*Figure 3 - Distributed quantum computer architecture. Source: elaborated by the author (2022).*

We see here that its modular representation allows the addition of new future layers, such as one specially dedicated to communication protocols and integration between processing cores, whether in a multicore machine or interaction with network infrastructure and communication protocols. We highlight three layers of functionality grouping:

• **Programming layer:** *where we find the programming resources and framework libraries available on the target platform. Currently, each platform vendor offers a distinct set*

*of features aimed at the developer, creating a wide variety of programming, simulation, and results from visualization tools.*

*For this research, we are using IBM Quantum platform and quantum providers from MICROSOFT AZURE QUANTUM, with machines from RIGETTI, IONQ and QUANTINUUM, through framework as QISKIT development kit and AZURE Quantum Dev Kit [130] [131], with tools for resources estimation and execution in a simulated environment and submission to real quantum computers accessible over the internet, through free test subscriptions;*

- ***Compilation layer:*** *this layer covers most of the actions for preparing and translating quantum circuits intended for execution in a processing schedule. Here, we see the steps of scheduling, optimization, decomposition, topological mapping, and logical partitioning of circuits. The present research proposes to add to this group the distribution stage, which deals with the physical partitioning scenarios of quantum circuits for machines with multiple processing cores and networks connecting multiple machines. Additional layers intended for the support of network protocols and connection negotiation can be considered outside this group due to the modular nature of this proposed architecture;*

- ***Execution layer:*** *finally, the execution layer receives the quantum circuit closer to the implemented physical qubits. Here, we see the decomposition into wave control instructions or native circuits of the technology adopted to build the target quantum computer. Different technologies are available, and there is no consensus on the most promising one for the next stage of quantum computing; for this research, we will use hardware providers in different technologies for qubits, with resources from MICROSOFT Azure Quantum platform and IBM Quantum platform. In the context of a distributed quantum computing architecture, the execution layer would be responsible for coordinating and submitting physical partitions of circuits to different agents involved in the distributed system.*

With this modular approach, we can explore the different treatment steps of a quantum circuit and highlight the processes that benefit the execution context in distributed scenarios.

In this context, a distributed quantum circuit (DQC – Distributed Quantum Circuit), also called a quantum multicomputer [21], consists of quantum circuits of limited capacity in k partitions that are physically distant from each other and emulate the functionality of a large quantum circuit. The partitions of a distributed environment communicate through communication qubits dedicated to entangled links, being considered an expensive resource for current NISQ computers.

Thus, the distribution of circuits in a distributed machine architecture involves the physical partitioning of the circuit for its encapsulation and submission to different machines. Likewise, important considerations must be made about the protocols and communication primitives used in communicating quantum information between machines and mechanisms for synchronization and coordination of results.

**5 – Challenges for distributed quantum computing**

We have seen so far that quantum algorithms are usually expressed in the form of quantum circuits, which are usually drawn as circuit diagrams in the literature. The big challenge for more sophisticated algorithms is that the representation circuit becomes very large and complicated to handle and read. Especially for computers with few resources, with dozens of qubits, for example, the challenge is even more significant due to decoherence or relaxation errors during circuit processing, which for more significant scenarios, causes the loss of quantum information and, consequently, the generation of invalid results.

For distributed quantum computing, the scenario becomes even more challenging since complex circuits have an infinite of gates of multiple qubits, representing potential communication links between distributed subcircuits, which demands an expensive communication resource through links entangled between the processing units involved [3].

The depth of a circuit is also a measure of quantum program performance. GYONGYOSI [128] defines the depth of a quantum circuit as the number of time steps (or time complexity) needed to perform the quantum operations. For circuits of more significant depth, a greater time complexity is observed, and emerging errors may occur, generating the loss of quantum information throughout the process. Decoherence is more relevant in the context of NISQ machines, and because of that, Quantum Circuit Optimization (QCO) is a relevant topic for research, considering the search for circuits with lower depth [129] and consequent lower complexity of time.

We can relate distributed quantum computing with many current challenges, which still open space for intense research. Among the main ones, we mention:

- *Optimization of quantum circuits by pattern matching;*
- *Circuit depth reduction;*
- *Translation of operations to the basic set of instructions supported by the target QPU architecture through optimized transpilation;*
- *Grouping of gates and operations by equivalence;*
- *Analysis of circuits by dependency relationship between operations;*
- *Anticipation of operations by analysis of persistence time in qubits;*
- *Analysis of constraints and resources of the communication link infrastructure in use.*

Numerous previous works have addressed these challenges with an eye on the realization of distributed quantum computing. Below we quote some critical works that support the future activities of this research.

FENG [22] presented an algebraic language for the representation of circuits, allowing the construction and manipulation of algorithms in a more optimized and fluid way. At the time, circuit diagrams were the primary study tool, with the challenges above for more complex algorithms. The approach brought a necessary rigor and formalism to distributed quantum computing and quantum protocol analysis in later years.

YOUNIS and UANCU [51] presented a circuit optimization algorithm that produced circuits on average 13% smaller compared to other optimizers. The approach also produced circuits with a 12% reduction in the number of 2-qubit gates used, which is a good metric for the communication context or relationship between subcircuits for multicore scenarios.

LIU et al. [86] presented a formal circuit reduction approach through Hamiltonian Theory, using a series of transformations in the form of a linear combination of operations. The technique is applied in the context of simulation of electronic structures, indicated by the author as a potential application of great interest for quantum computing. The approach was tested by studying the potential energy curve of the $H_2O$ (water) molecule, with better chemical precision achieved with a much smaller circuit depth. However, the procedure was applied only to VQE – Variational Quantum Eigensolver algorithms indicated for the study of the simulation of electronic structures. Due to its hybrid nature, VQE is executed in two phases, the first in quantum devices for the generation of estimates and preparation of states, followed by variational optimization on classical machinesidentified two circuit optimization rules, labeled for optimization of the solution parameters [92].

FÖSEL [88,89] developed a specific approach for circuit optimization using reinforcement machine learning algorithms, with an average circuit depth reduction of 27% and gate count reduction of 15%, for benchmark circuits of 12 randomly generated qubits. The agent trained with 12 qubits was then applied in the QAOA circuit - Quantum Approximate Optimization Algorithm of optimization with 50 qubits. A batch of 100,000 transformations took one week to process. The presented transformation technique is based on the removal of self-canceling operations or the reversal of self-switched operations. The authors identified two circuit optimization rules, labeled *soft* and *hard*: *hard* optimizations generate circuit depth reduction, with the removal of canceling operations, with no impact on solution convergence; the *soft* optimizations are those that rearrange the quantum gates, according to the adjacent operations. This same classification between *soft* and *hard* optimizations will be used in the present research, focusing on optimizing and grouping communication gates between partitions in the context of distributed circuits.

HOUSHMAND [90] et al. present a specialized genetic algorithm (GA - Genetic Algorithm) for the optimization of circuits, with a special focus on the reduction of communication qubits, evolving the previous work of the authors [31].

Among the circuit optimization and reduction techniques, we also find the partitioning of quantum circuits by time. As introduced by BAKER et al. [93], an input quantum circuit can be partitioned into blocks that group operations and qubits into so-called clusters, considering dependency relationships and execution schedules, which highlights possible situations of optimization and grouping of operations. This approach aims to keep qubits and operations closely related in circuit clusters, reducing the cost of communication between distributed subcircuits. The partitioning referenced by the author is linked to the vertical cut of the circuit, which is also interesting for optimizing circuits in the compilation process. For the scenario of horizontal circuit partitioning, supporting the distribution in processing agents, the author uses the KL algorithm [13], which will be discussed in future chapters.

### 6 – Quantum Teletransportation

We can consider a quantum processor as a quantum processing unit or QPU – Quantum Processor Unit. An alternative for greater scalability with NISQ machines is the use of a multi-QPU or multicore machine approach [7]. Current hardware technologies offer specific challenges for realizing multicore machines, such as the cost of communication between QPUs.

In this context, communication between QPUs must occur through quantum teleport protocols, as proposed by BENNETT et al.[40][123]. Quantum teleportation is accomplished by creating a pair of entangled qubits (ebits – entanglement qubits), which share a Bell state. When information is sent from one QPU to another, the ebit is consumed, and the sent qubit state becomes unavailable (Non-Cloning Theory). The cost of producing ebits between QPUs is high, being targeted for reduction for the distribution of more efficient quantum circuits.

Thus, communication between partitions through quantum teleportation protocols consumes dedicated qubits for communication, for example, implementing non-local operators such as CNOT or CZ. The following figures illustrate the representation of these communication links, with dedicated qubits for communication:

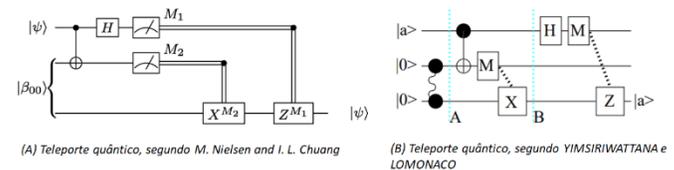

(A) Teleporte quântico, segundo M. Nielsen and I. L. Chuang

(B) Teleporte quântico, segundo YIMSIRIWATTANA e LOMONACO

*Figure 4 – Shows quantum teleportation consuming communication qubits, according to NIELSEN [73] and LOMONACO [3].*

In the present work, we adopted the BACK and FORTH (bidirectional) approach of LOMONACO [3] with the primitives of CAT-ENTANGLER and CAT-DISENTANGLER, which consume two ebits of communication.

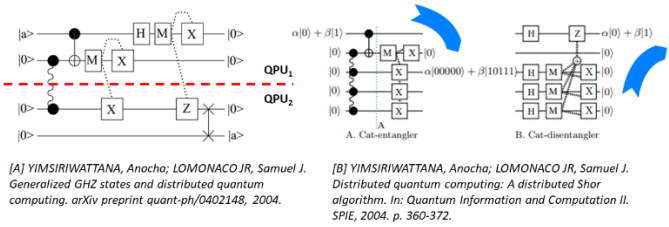

[A] YIMSIRIWATTANA, Anocha; LOMONACO JR, Samuel J. Generalized GHZ states and distributed quantum computing. arXiv preprint quant-ph/0402148, 2004.

[B] YIMSIRIWATTANA, Anocha; LOMONACO JR, Samuel J. Distributed quantum computing: A distributed Shor algorithm. In: Quantum Information and Computation II. SPIE, 2004. p. 360-372.

Figure 5 – cat-entangler and cat-disentangler as created by LOMONACO et al. Source: [3].

As described by LOMONACO, we can identify reusable qubits for remote operations, as illustrated in the previous figure. Note that the control qubit can act on the composite operator U, which acts on the operations $U_2$, $U_1$, and CNOT in a non-local way. This behavior can be expanded and recognized in scenarios with similar operations nearby, such as CX (CONTROLLED-X), CZ (CONTROLLED-Z), and CP (CONTROLLED-PHASE), allowing the grouping of gates for optimization in the circuit segmentation process.

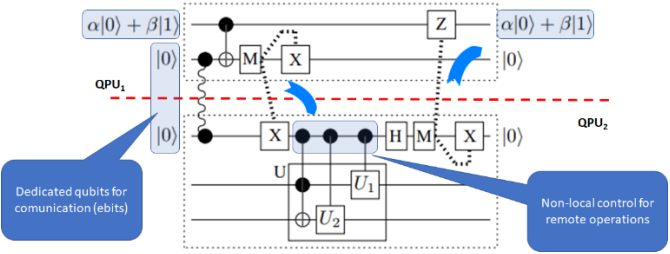

Figure 6 - Quantum circuit partitioning scenario between 2 QPUs, if the operation U = CNOT.$U_1$.$U_2$. Source: LOMONACO [3].

Thus, the partitioning of a quantum circuit for two QPUs, for example, can be performed considering binary gates as NON-LOCAL controls between partitions. Cuts made in the circuit will generate consumption of communication qubits, thus being the reduction target due to the associated cost of generation and impact on system performance.

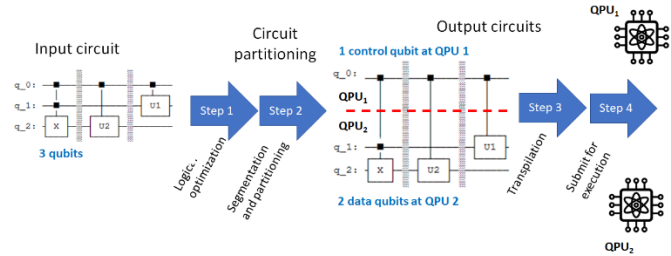

Figure 7 - Quantum circuit partitioning scenario between two QPUs. Source: designed by the author.

### 7 – Hypergraphic representation of quantum circuit

In the literature, we find different approaches for the decomposition and fragmentation of quantum circuits to treat this problem.

One of these approaches uses the hypergraphic representation of the quantum circuit with the application of heuristics to solve the problem.

The partitioning problem of hypergraphs is considered "NP-hard" equivalent to the partitioning problem of quantum circuits.

As suggested in previous works in the literature, a hypergraphic representation for quantum circuits follows the definitions below:

| hypergraphic partitioning | Hypergraphic partitioning of a quantum circuit |
|---|---|
| Vertices or nodes | Qubits |
| Edges | Non-local control ports such as CNOT, CZ, TOFFOLI, CCZ |
| Sub-partition of the hypergraph | quantum circuit partition |
| Minimal edge cutting (mincut) | Minimum number of consumed ebits |
| Balance of vertices between sub-partitions | Balancing qubits between QPUs of a distributed system. |

Figure 8 – Relationship between hypergraphic representation of quantum circuits and components of a classical hypergraphic approach. Source: elaborated by the author (2022).

To translate the quantum circuit representation into hypergraphic representations, we use the following pseudo-code, the gate from the circuit, to build vertices and edges in the hypergraph. Another critical difference is the consolidation of other statistics in this process, saving lists of nodes, gates, directions, and vertices that can be used in future steps in this method.

**Algorithm 1 – Translating quantum circuits into hypergraph representation.**

```
1   input: circuit
2   output: (H, G, V, E)
3   begin
4       G ← ∅           #quantum gates
5       V ← ∅           #vertices representing qubits
6       E ← ∅           #edges representing binary gates
7       H ← ∅           #hypergraph
8       foreach gate in circuit do
9           G ← G ∪ { gate }
10          V ← V ∪ { vertices_from( gate ) }
11          if isCZ (gate) or isCX (gate) then
12              E ← E ∪ { edge_from( gate ) }
13          H ← H ∪ { G, V, E }
14  end
```

A classic approach to the hypergraph partitioning problem is given by the heuristic of **Charles Fiduccia** and **Robert Mattheyses** [10], known as the FM algorithm, or Fiduccia-Mattheyses.

- *Hypergraphic partitioning in a heuristic and iterative way.*

- *Much faster than the Kernighan-Lin method (KL), which comes with big-O($n^3$), whereas FM offers big-O(n).*
- *FM handles one vertice at a time, while KL handles two vertices simultaneously.*
- *FM can be easily implemented through a doubly linked list, allowing for variations in its structure and mode of iterations.*

A pseudocode for the FM algorithm, with bipartite iterative execution, follows below, as illustrated by GOTTESBÜREN [127]:

**Algorithm 2 – Steps from Fiduccia-Mattheyses, as described by GOTTESBÜREN**

```
1   Create initial partitioning;
2   While cutsize is reduced {
3       Unlock all nodes;
4       While valid moves exist {
5           Use bucket data structures to find unlocked node in each partition that most
6               improves cutsize when moved to other partition;
7           Move whichever of the two nodes most improves cutsize while not exceeding
8               partition size bounds;
9           Lock moved node;
10          Update nets connected to moved nodes, and nodes connected to these nets;
11      } endwhile;
12      Backtrack to the point with minimum cutsize in move series just completed;
13  } endwhile;
```

Using this method, we can translate a quantum circuit into a hypergraphic representation, run the FM algorithm and then translate the generated partitions into new subcircuits for distribution.

The following figure shows an example of a circuit with six qubits, translated into hypergraphic representation and then bipartite by the classical FM method, as used in this work.

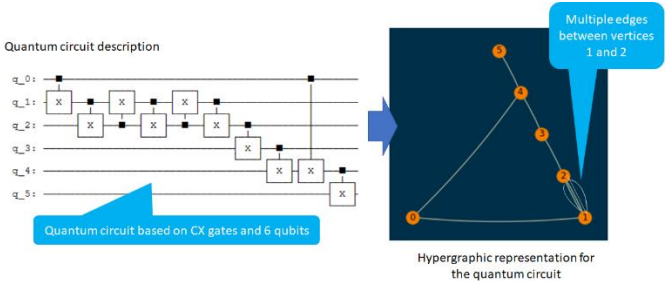

*Figure 9 – Translating a quantum circuit into hypergraphic representation.*

A random partitioning without vertices grouping can generate a segmentation with high communication cost and consumption of ebits. LOMONACO [3] illustrated the reuse of ebits for control qubits on non-local gates. MARTINEZ [27] suggested the creation of grouping vertices for sequences of 3 CZ gates in hypergraphic representation, indicating reuse. CX and CZ port groupings were identified according to the following patterns:

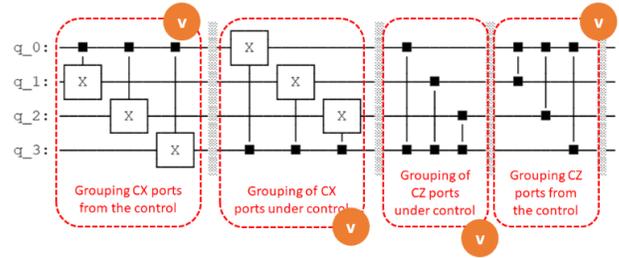

*Figure 10 – Grouping patterns for CX (CONTROLLED-X) and CZ (CONTROLLED-Z) gates to reuse ebits during circuit partitioning.*

Likewise, CP gates with angles with the same phase settings will be grouped:

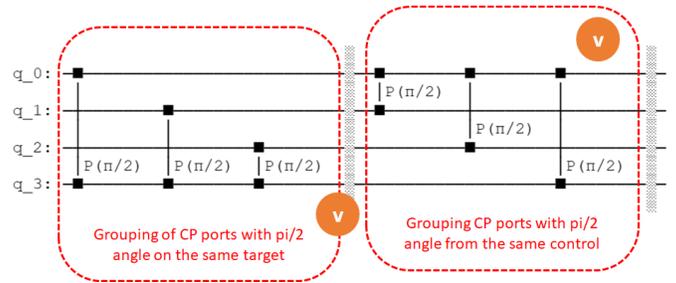

*Figure 11 – Grouping patterns for CP (CONTROLLED PHASE ROTATION) to reuse ebits during circuit partitioning.*

In the end, generating clustering patterns allows the creation of additional vertices in the hypergraph, reducing cutoff points and optimizing communication qubits. A cluster vertice generates a single cat-entangler and cat-disentangler pair among the generated partitions for distribution.

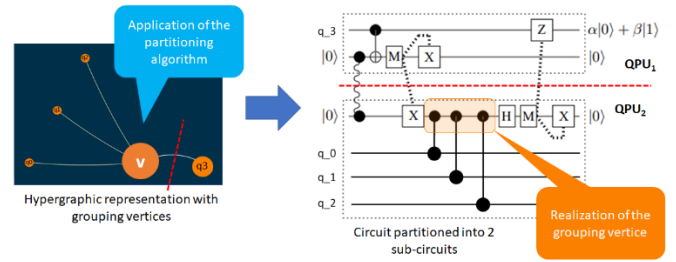

*Figure 12 – Circuit partitioning process, with the consumption of 1 pair CAT-ENTANGLER and CAT-DISENTANGLER per grouping vertice in the circuit.*

### 7 – Experiments and Examples

To explore the impact of quantum circuit partitioning in hypergraphic representation, we run several experiments with quantum circuits in different configurations, comparing dimensions and results for several partitions created.

**7.1 – Generating Random Circuits –** for this research stage, we are using random circuits generated via the Munich Quantum Toolkit Benchmark Library https://www.cda.cit.tum.de/mqtbench/.

**MQT Bench** is a quantum software tool and benchmark suite which offers the same benchmark algorithms on different levels of abstractions. For our experiments, we use the following circuits in different settings for optimization and many qubits:

- Amplitude Estimation (AE)
- Quantum Fourier Transformation (QFT)
- Greenberger-Horne-Zeilinger (GHZ)
- Variational Quantum Eigensolver (VQE)

**7.2 – Running random bipartition –** for the set of benchmark algorithms, we got the variables from each circuit as the number of qubits, the number of binary gates (CX, CZ, and CP), width, depth, and the OpenQASM representation. As a first experiment, we made random partitionings, cutting the circuit with an equal number of qubits between the two partitions. In the end, we calculated the number of ebits used in this process, against the original number of binary gates and potential cuts, before the partitioning.

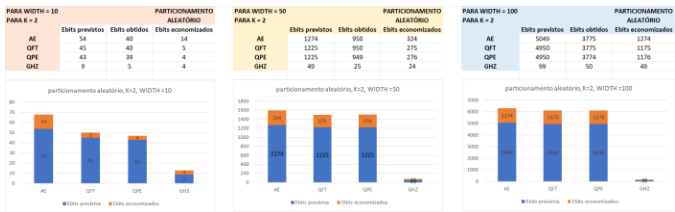

*Figure 13 – Results of partitioning using the random method over the benchmark circuits and for two processing units.*

**7.3. Running FM bipartition –** as a second experiment, we applied the classical FM heuristic over the benchmark circuits with a bipartite approach, resulting in better results of using ebits when compared to the random method.

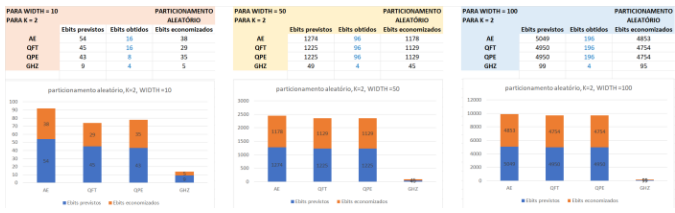

*Figure 14 – Results of partitioning using the classical FM method over the benchmark circuits.*

**7.4. Classical FM method with a pattern of reuse in hypergraph –** finally, still in the scenario of the bipartition approach, we are applying the classical FM method to the set of hypergraph representations, adding vertices for grouping patterns, reusing ebits for nearby gates as CX, CZ and CP with same settings. We expect to consolidade the telemetry from this experiment, to obtain better results compared to random partitioning and classical FM method without grouping vertices.

<div align="center">

**8 – Conclusions**

</div>

With the partitioning of quantum circuits in the form of hypergraphs, we observe that:

- *The bipartite recursive FM algorithm is faster than the algorithm KL with the same approach, being more efficient than the random cut;*
- *The bipartite partitioning performed with the classic FM method improved by more than 50% against partitioning performed randomly on the same AE, QFT, and QPE calibration circuits. GHZ circuits with 10, 50, and 100 qubits showed improvements between the order of 46% against random partitioning.*
- *The segmentation of the input circuit is necessary for circuits of great depth, creating segments that will be targets of partitioning by stages.*
- *The grouping of gates in additional grouping vertices in the hypergraph generates clear-cut points in the circuit, facilitating the indication of communication primitives such as CAT-ENTANGLER and CAT-DISENTANGLER in the stages of reconstruction of the quantum circuit.*

This research points to generating circuit development patterns, better segmentation, and reusing groups of ebits to generate lower communication costs in distributed systems.

<div align="center">

**8 – Future works**

</div>

As future activities, we must consolidate data for partitioning over hypergraphs with clustering vertices CX, CZ, and CP, following different combinations of gate patterns. Partitioning performed with the classical FM method on clustering vertices should obtain better results than previous experiments.

Other families of quantum circuits are planned to be used in this exploration, helping the research to discover new insights about the impact of dimensions and grouping gates before the partitioning process.

Still, we must consider a variation on the Fiduccia-Mattheyses algorithm, supporting a direct K-WAY approach for K > 2 partitions. In these experiments, we must also consider additional gate clustering vertices on the same target qubits, observing the results obtained.